\documentclass[conference]{IEEEtran}
\usepackage{myStyleIEEE}
\usepackage{ushort}
\IEEEoverridecommandlockouts

\usepackage{siunitx}
\AtBeginDocument{\DeclareSIUnit{\MWh}{MWh}}
\AtBeginDocument{\DeclareSIUnit{\kWh}{kWh}}
\AtBeginDocument{\DeclareSIUnit{\voltampere}{VA}}

\usepackage{amsmath}
\usepackage{mathtools}
\usepackage{isomath}
\let\underbrace\LaTeXunderbrace

\usepackage{eurosym}
\DeclareRobustCommand{\officialeuro}{%
  \ifmmode\expandafter\text\fi
  {\fontencoding{U}\fontfamily{eurosym}\selectfont e}}

\usepackage{acronym}
\usepackage[]{optidef}
\usepgfplotslibrary{groupplots}
\definecolor{darkgreen}{rgb}{0.0, 0.5, 0.0}

\begin{document}

\acrodef{RES}{Renewable Energy Source}
\acrodef{SG}{Synchronous Generator}
\acrodef{VSC}{Voltage Source Converter}
\acrodef{RoCoF}{Rate of Change of Frequency}
\acrodef{DAE}{Differential Algebraic Equation}

\title{Dynamic Optimization of Virtual Inertia and Damping in Converter-Based Power Systems}
\renewcommand{\theenumi}{\alph{enumi}}

\newcommand{\jovan}[1]{\textcolor{magenta}{$\xrightarrow[]{\text{J}}$ #1}}
\newcommand{\maitraya}[1]{\textcolor{blue}{$\xrightarrow[]{\text{M}}$ #1}}
\newcommand{\ognjen}[1]{\textcolor{red}{$\xrightarrow[]{\text{O}}$ #1}}

\author{
\IEEEauthorblockN{Jovan Krajacic\IEEEauthorrefmark{1}, Maitraya Avadhut Desai\IEEEauthorrefmark{1}, Ognjen Stanojev\IEEEauthorrefmark{2}, Gabriela Hug\IEEEauthorrefmark{1}}%
\IEEEauthorblockA{\IEEEauthorrefmark{1} EEH - Power Systems Laboratory, ETH Zurich, Switzerland} %
\IEEEauthorblockA{\IEEEauthorrefmark{2} ABB Corporate Research Center, Switzerland} %
Emails: jkrajacic@ethz.ch, \{desai, hug\}@eeh.ee.ethz.ch, ognjen.stanojev@ch.abb.com
\thanks{Research supported by NCCR Automation, a National Centre of Competence in Research, funded by the Swiss National Science Foundation (grant number 51NF40\_180545).}
}

\maketitle
\IEEEpeerreviewmaketitle

\begin{abstract}
The transition towards a sustainable power system is enabled by the replacement of conventional synchronous generators with converter-interfaced renewable energy sources. However, the resulting loss of rotational inertia and governor damping causes significant frequency deviations and can therefore cause instability. The focus of this paper is the optimal allocation of virtual inertia and damping in the power system activated by established converter control schemes. To this end, we propose a novel dynamic optimization algorithm that considers performance metrics for system stability, cost-efficiency, and resilience. In addition, our algorithm considers the magnitudes and locations of disturbances in the power system for the optimal allocation. Finally, we validate our approach on a three-area system and also compare our results with a $\mathcal{H}_2$ system-norm-based allocation approach.
\end{abstract}

\begin{IEEEkeywords}
dynamic optimization, grid-forming, virtual inertia, voltage source converters
\end{IEEEkeywords}

\section{Introduction} \label{sec:intro}



The traditional design and operation of power systems are undergoing a significant transformation, shifting towards converter-interfaced generation driven by the integration of \acfp{RES}. Consequently, \acfp{SG} found in conventional power plants are being decommissioned, together with their inherent electromechanical characteristics -- rotational inertia and damping. This development leads to an increased number of system frequency excursions and stability incidents in general, as converter-interfaced \acp{RES} do not fundamentally offer the stabilizing response conventional \acsp{SG} provide \cite{TIELENS2016999}. To reinforce the stability of the weakened system, the concept of virtual synchronous generator (machine) control \cite{Bevrani2014} was introduced for grid-connected \acfp{VSC}, which enables converter-interfaced \acp{RES} to emulate the inertial and damping behavior of \acp{SG}.


In recent years, extensive research has been conducted on virtual synchronous generators, focusing on control design \cite{Bevrani2014,Hesse2007}, small-signal stability analysis \cite{DArco2014,Markovic2021}, fault ride-through \cite{Heng2020}, transient stability analysis \cite{Cheng2020}, etc. Nevertheless, most of the research has been directed toward device-level aspects, and the subsequent step of determining the optimal placement and sizing of virtual inertia and damping across the system remains an ongoing research topic. While the total amount of system inertia is relevant to ensure grid resilience \cite{ULBIG20147290}, it was demonstrated in \cite{poolla2017optimal} that its spatial distribution is also critical.


Two groups of methods have been proposed thus far for determining the optimal placement and sizing of virtual inertia and damping: (i) eigenvalue optimization-based methods \cite{BorscheLiuHill,Uros} and (ii) $\mathcal{H}_2$~norm minimization methods \cite{poolla2017optimal,dfig,Ademola2018, 8607122,IREP2017}. The algorithms proposed in \cite{BorscheLiuHill,Uros} iteratively adjust virtual inertia and damping parameters to optimize damping of modes of a linearized system model. Although these methods allow for detailed differential-algebraic power system models to be used and result in high-quality solutions, the overall procedure is computationally demanding and no convergence guarantees can be given. In contrast, the methods in \cite{poolla2017optimal}, \cite{dfig,Ademola2018, 8607122, IREP2017} use network coherency as the performance metric evaluated on simplified and linear power system models. Specifically, the $\mathcal{H}_2$ system norm is selected as the objective function to minimize the impact on system states of stochastic fluctuations, such as those caused by loads and \ac{RES}, as well as impulsive event-like faults. The proposed algorithms are scalable, computationally efficient, and result in improved system resilience.

Despite the apparent advantages outlined above, minimizing the $\mathcal{H}_2$~norm presents certain limitations. Notably, the algorithms do not directly constrain state trajectories post-disturbance, which in the case of larger disturbances could trigger under-frequency protection schemes due to higher \ac{RoCoF} and low frequency nadir values. Moreover, step disturbances are more prevalent in power systems since they better reflect load variations and generation outages, making the $\mathcal{H}_2$~norm's impulse-disturbance-based allocation applicable only to specific scenarios, such as faults. In addition, virtual inertia and damping relate to certain operational and design costs (energy delivery and converter sizing). Although the $\mathcal{H}_2$ norm limits the budget for available virtual inertia and damping at \ac{RES} buses, it does not optimize the cost of providing these services. Rather, the $\mathcal{H}_2$ norm focuses solely on frequency stability, while the economic aspect of the optimization problem is neglected. Finally, the relevant studies in \cite{poolla2017optimal}, \cite{dfig} neglect the virtual damping placement.

Considering the limitations of the methods proposed in the literature so far, this paper proposes an optimal virtual inertia and damping allocation algorithm based on dynamic optimization, which integrates the benefits of minimizing the $\mathcal{H}_2$~norm while also reducing the virtual inertia and damping cost at each \ac{RES} bus, thereby including the economical aspect in the optimization problem. Furthermore, the post-disturbance trajectories are directly bounded in the constraints of the optimization problem to limit the \ac{RoCoF} and frequency nadir, as opposed to the approach of constraining analytically-derived expressions for the aforementioned metrics~\cite{chu2024schedulingsoftwaredefinedmicrogridsoptimal}. Considering that different disturbance locations and magnitudes might lead to different optimal inertia and damping allocations, we use the probability of disturbance magnitudes and locations in combination with the expected value formula to determine the recommended inertia and damping allocation values for all \ac{RES} buses. Finally, the performance of the proposed approach is evaluated on a standard 3-area system.

The subsequent sections of this paper are organized as follows. Section~\ref{sec:modeling} provides a brief overview of virtual inertia, damping concepts, and the power system model used. In Section~\ref{sec:dyn_opt}, we introduce the proposed dynamic optimization problem. Section~\ref{sec:res} presents the conducted case studies to assess the optimization performance, while Section~\ref{sec:concl} concludes the paper.

\section{System Modeling} \label{sec:modeling}

\subsection{Virtual Inertia and Damping} \label{sec:hnet}
\begin{figure}
    \centering
    \includegraphics[scale=0.95]{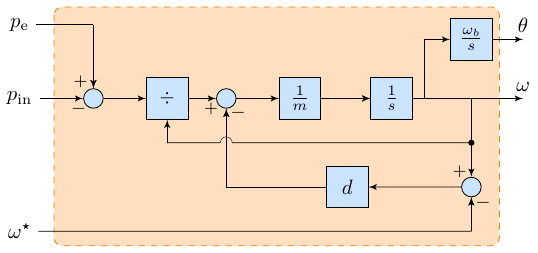}
    \vspace{-0.2cm}
    \caption{Control block diagram of virtual synchronous machine active power controller. Constants $m$ and $d$ represent the virtual inertia and damping.}
    \label{fig:VSM_control}
    \vspace{-0.5cm}
\end{figure}

As previously discussed, the virtual synchronous machine control developed for \acp{VSC} emulates virtual inertia and damping by adjusting the converter frequency in response to the difference between the power setpoint $p_\mathrm{in}$ and the measured output power $p_\mathrm{e}$. The control blocks typically implemented in virtual synchronous machine active power control are presented in Fig.~\ref{fig:VSM_control}. The equations representing the controller response are given in the time domain and per unit by
\begin{align}
    m\ddt{\omega} &= \frac{1}{\omega} (p_\mathrm{in}-p_\mathrm{e}) - d(\omega-\omega^\star), \label{eq:vsm_nonlinear_1}\\
    \ddt{\theta} &= \omega_{\rm b}\omega,\label{eq:vsm_nonlinear_2}
\end{align}
where $\omega$ is the converter frequency, $\theta$ is the converter angle, $m$ and $d$ are the virtual inertia and damping parameters, respectively. Furthermore, $\omega^\star$ is the frequency setpoint and $\omega_{\rm b}$ is the base system frequency. We further linearize \eqref{eq:vsm_nonlinear_1} around a steady state and then substitute \eqref{eq:vsm_nonlinear_1} into \eqref{eq:vsm_nonlinear_2} considering $m^{\prime}=m/\omega_{\rm b}$ and $d^{\prime}=d/\omega_{\rm b}$ to obtain
\begin{equation}\label{eq:vsm_linear}
    m^{\prime}\ddttwo{\theta}+d^{\prime}\ddt{\theta} = p_\mathrm{in}-p_\mathrm{e}.
\end{equation}
It can be shown that under certain parameter selection criteria and the above linearization assumptions, the considered virtual synchronous machine control is equivalent to grid-forming droop control \cite{Ofir2018}. Furthermore, we can observe that the derived linear form \eqref{eq:vsm_linear} also corresponds to the swing equation model of synchronous machines. However, while synchronous generators have constant $m$ and $d$ parameters determined by the electromechanical properties of the generator, these parameters can be freely tuned in the case of grid-connected converters controlled as virtual synchronous generators. 

\subsection{Network Model} \label{subsec:network}
We consider a power network modeled by a connected graph $\mathcal{G}(\mathcal{N},\mathcal{E})$, with $\mathcal{N} \coloneqq \{1,\dots,n\}$ denoting the set of nodes, and $\mathcal{E} \subseteq \mathcal{N}\times\mathcal{N}$ representing the set of network lines. Using Kron reduction, passive loads and nodes without injection are reduced such that only $n$ active buses remain, which are further partitioned as $\mathcal{N} = \mathcal{N}_\mathrm{sg} \cup \mathcal{N}_\mathrm{c}$ to separate sets of nodes hosting synchronous generators and converters providing virtual inertia and damping. Nodal dynamics for all $i\in\mathcal{N}$ are given by
\begin{equation}
    m_i^{\prime}\ddttwo{\theta_i} + d_i^{\prime}\ddt{\theta_i} = p_{\mathrm{in},i} - p_{\mathrm{e},i},
    \label{eq:swing}
\end{equation}
which corresponds to \eqref{eq:vsm_linear} in case $i\in\mathcal{N}_\mathrm{c}$. On the other hand, if $i\in\mathcal{N}_\mathrm{sg}$, then \eqref{eq:swing} represents the swing dynamics for the generator rotor angle $\theta_i$, where $m_i^{\prime}$ denotes the generator rotational inertia and $d_i^{\prime}$ accounts for the generator frequency damping, with $m_i^{\prime}>0$ and $d_i^{\prime}>0$. Additionally, $p_{\mathrm{in},i}$ refers to the mechanical input power and $p_{\mathrm{e},i}$ refers to the electrical output power. In general, each bus $i$ could host an ensemble of devices, and the quantities $m_i^{\prime}$ and $d_i^{\prime}$ represent lumped parameterizations of the aggregate behavior.

The electrical output power at each bus can be calculated using the DC power flow approximation, assuming identical unit voltage magnitudes at each bus and purely inductive lines, i.e.
\begin{equation}
    p_{\mathrm{e},i} = \sum_{j=1}^{n} b_{ij} (\theta_i - \theta_j),
    \label{eq:dc}
\end{equation}
where $b_{ij}>0$ is the susceptance between the nodes $i$ and $j$. 
Considering the relation between angle and frequency \eqref{eq:vsm_nonlinear_2}, and stacking \eqref{eq:swing} and \eqref{eq:dc} for all nodes, the power system dynamics can be written in the state space form as
\begin{equation}
\underbrace{
\ddt{
\begin{bmatrix}
    {\boldsymbol{\theta}} \\
    {\boldsymbol{\omega}} 
\end{bmatrix}}}_{\dot{\boldsymbol{x}}}
    = 
\underbrace{
\begin{bmatrix}
\mathbf{0} & \omega_{\rm b}\mathbf{I} \\
-\mathbf{M}^{-1}\mathbf{L} & -\mathbf{M}^{-1}\mathbf{D}
\end{bmatrix}}_{\mathbf{A}}
\underbrace{
\begin{bmatrix}
    {\boldsymbol{\theta}} \\
    {\boldsymbol{\omega}}
\end{bmatrix}}_{\boldsymbol{x}}
    +
\underbrace{
\begin{bmatrix}
0 \\
\mathbf{M}^{-1}
\end{bmatrix}}_\mathbf{B}
\boldsymbol{p}_{\mathrm{in}}.
\label{eq:state}
\end{equation}
Here, the states $\boldsymbol{\theta}$ and $\boldsymbol{\omega}$ represent stacked vectors of the angular positions and velocities and $\boldsymbol{p}_{\mathrm{in}}$ stacks $p_{\mathrm{in},i},\forall i\in\mathcal{N}$. Matrices $\mathbf{M}=\mathrm{diag}(\{m_i^{\prime}\}_{i=1}^n)$ and $\mathbf{D}=\mathrm{diag}(\{d_i^{\prime}\}_{i=1}^n)$, whereas $\mathbf{L}\in\mathbb{R}^{n \times n}$ is the Laplacian (susceptance) matrix of the considered power network. The diagonal elements of the Laplacian matrix are $l_{ii}=\sum_{j=1, j \neq i}^{n} b_{ij}$, whereas the off-diagonal elements are $l_{ij}=-b_{ij}$. Note that $\mathbf{L}$ is symmetric.

\section{Methodology} \label{sec:dyn_opt}

\subsection{Dynamic Optimization Based Optimal Allocation}\label{subsec:dyn_opt}

We aim to optimally allocate virtual inertia and damping in the system, while ensuring that certain metrics are within the prescribed thresholds. These metrics are the frequency nadir and \ac{RoCoF}. The frequency nadir is the lowest frequency reached after a disturbance and \ac{RoCoF} is the maximum rate at which the frequency changes. Both metrics are derived from the state trajectories for the system described in Section~\ref{sec:modeling}. Thus, we formulate the optimal allocation problem considering the state-space formulation of the system in the constraints. Such optimization problems that are constrained by \acp{DAE} are called dynamic optimization problems \cite{Biegler_book}.

Considering the described system, we optimize the inertia and damping at the \ac{VSC} buses $m_j,\,d_j,\,\forall j \in \mathcal{N}_{\rm c}$, while the \ac{SG} buses $k\in\mathcal{N}_{\rm sg}$ have fixed parameters. The optimization is performed over the time interval $t\in[t_0,t_\mathrm{f}]$ where a fault is simulated at $t=t_0$ through the modification of the vector $\vect{p_{\rm{in}}}$. The optimization problem is given as
\begin{subequations}\label{eq:md_dyn_opt}
\begin{align}
    \underset{m_j,d_j,\vect{x}}{\text{min}} \quad & C_\kappa \left(\,\sum_{j\in\mathcal{N}_{\mathrm{c}}}\left(c_{{\rm{m}}_j}m_j+c_{{\rm{d}}_j}d_j\right)\right) +C_{\vect{x}}\left( \dot{\boldsymbol{x}}^{\top} \dot{\boldsymbol{x}}\right)\nonumber\\ &\qquad\qquad\qquad\quad + C_\xi\left(\sum_{i\in\mathcal{N}} (\xi_{\dot\theta,i}^2+\xi_{\dot\omega,i}^2)\right) \label{eq:ObjFun} \\
    \text{s.t.} \quad & \forall i \in \mathcal{N},\,\forall j \in \mathcal{N}_{\mathrm{c}},\,\forall t \in [t_0,t_{\mathrm f}],\nonumber \\
    & \dot{\vect{x}} = \mat{A}\vect{x}+\mat{B}\vect{p_{\mathrm{in}}}, \label{eq:DAEconstraint} \\
    & \vect{x}(t_0)=\vect{x}_0,  \label{eq:init_cond_constraint} \\
    & {\dot{\theta}}_{\mathrm {min}}-\xi_{\dot{\theta}_i} \leq \dot\theta_{i} \leq {\dot{\theta}}_{\mathrm {max}}+\xi_{\dot{\theta}_i},\label{eq:freq_constraint} \\
    & {\dot{\omega}}_{\mathrm {min}}-\xi_{\dot{\omega}_i} \leq \dot\omega_{i} \leq {\dot{\omega}}_{\mathrm {max}}+\xi_{\dot{\omega}_i},\label{eq:rocof_constraint} \\
    & \underline{m}_{j} \leq m_{j} \leq \overline{m}_{j}, \label{eq:inertia_constraint} \\
    & \underline{d}_{j} \leq d_{j} \leq \overline{d}_{j}, \label{eq:damping_constraint} \\
    & \sum_{j\in\mathcal{N}_{\mathrm{c}}}m_j \leq m_{\mathrm {budget}},\label{eq:totalinertiabudget} \\
    & \sum_{j\in\mathcal{N}_{\mathrm{c}}}d_j \leq d_{\mathrm {budget}},\label{eq:totaldampingbudget} \\
    & \xi_{\dot{\omega}_i},\xi_{\dot{\theta}_i} \geq 0.  \label{eq:slack_positivity}
\end{align}
\end{subequations}

The objective function \eqref{eq:ObjFun} consists of three terms. The first term penalizes the sum of the inertia $m_j$ and damping $d_j$ used at the converter buses $j\in\mathcal{N}_{\rm{c}}$ with cost factors $c_{{\rm{m}}_j}$ and $c_{{\rm{d}}_j}$, respectively. The second term penalizes the post-disturbance energy dissipation through the derivative of state vector $\vect{x}$, namely angular position and velocity. Finally, the third term penalizes the slack variables used in the constraints imposed on the derivative of the states, i.e., \eqref{eq:freq_constraint} and \eqref{eq:rocof_constraint}. The tunable costs $C_{\kappa}$, $C_{\vect{x}}$, and $C_\xi$ adjust the weight of each objective function component. The coefficient $C_\xi$ is set significantly larger than $C_{\kappa}$ and $C_{\vect{x}}$, as constraint satisfaction is prioritized over minimizing the other terms. The ratio between $C_{\kappa}$ and $C_{\vect{x}}$ can be tuned to prioritize the cost of inertia and damping \emph{or} energy dissipation in the system.

The state-space model \eqref{eq:state} is used as a constraint in \eqref{eq:DAEconstraint} with the initial pre-fault condition specified in \eqref{eq:init_cond_constraint}. The fault for which the optimization is performed is specified in the vector $\vect{p_{\mathrm{in}}}$. For optimization problems with such \ac{DAE} constraints, the computational complexity of handling them is overcome by using discretization techniques. In particular, we employ orthogonal collocation over finite elements. This method first divides the continuous time horizon into finite elements and further into collocation points. Furthermore, the differential variables are approximated by polynomials defined using collocation points. We refer the reader to \cite[Chapter 10]{Biegler_book} for further details on orthogonal collocation. 

The constraints \eqref{eq:freq_constraint} and \eqref{eq:rocof_constraint} are imposed to maintain the frequency nadir and \ac{RoCoF} within their prescribed range $[\dot{\theta}_{\rm min},\dot{\theta}_{\rm max}]$ and $[\dot{\omega}_{\rm min},\dot{\omega}_{\rm max}]$, respectively. These constraints are softened with non-negative slack variables $\xi_{\dot\theta,i},\xi_{\dot\omega,i}$ \eqref{eq:slack_positivity} to ensure feasibility in case of overly restrictive ranges, limited virtual inertia and damping resources, and/or high disturbance magnitudes. The constraints \eqref{eq:inertia_constraint} and \eqref{eq:damping_constraint} limit the inertia and damping of each \ac{VSC} bus between lower and upper limits.  Finally, the total inertia and damping in the system are limited by budgets $m_{\rm{budget}}$ and $d_{\rm{budget}}$ in constraints \eqref{eq:totalinertiabudget} and \eqref{eq:totaldampingbudget}, respectively. Thus, the goal is to allocate inertia and damping to the buses based on necessity, prioritizing the allocation to the most affected buses. 

\subsection{Optimal Allocation Based on $\mathcal{H}_2$ norm  Minimization}\label{sec:H2norm_allocation}

The optimal allocation presented in \cite{poolla2017optimal}  tunes the inertia at all active buses while minimizing the squared $\mathcal{H}_2$ norm. For the $\mathcal{H}_2$ norm, the output performance vector is defined as
\begin{equation}
    \boldsymbol{y} = \underbrace{\begin{bmatrix}
            \mathbf{N}^{\frac{1}{2}} & 0 \\
            0 & \mathbf{S}^{\frac{1}{2}}
        \end{bmatrix}}_\mathbf{C}
        \begin{bmatrix}
            \boldsymbol{\theta} \\
            \boldsymbol{\omega}
        \end{bmatrix}, 
        \label{eq:h2_performance_output}
\end{equation}
where $\mathbf{N}$ is the Laplacian matrix of the system, and $\mathbf{S}=\mathbf{M}$ denotes the diagonal matrix of the inertia constants at each bus, as previously defined in Section~\ref{subsec:network}. Thus, the input-to-output map can be considered as $\mathcal{G}=(\mat{A},\mat{B},\mat{C})$. The squared $\mathcal{H}_2$ norm is therefore given as
\begin{equation}
     \|\mathcal{G}\|_2^2 = {\sum_{i=1}^{n} \, \int_{0}^{\infty} y_i(t)^{\top
     } y_i(t) \, \mathrm{d}t}.
     \label{norm}
\end{equation}
If unit impulse inputs are considered, the squared $\mathcal{H}_2$ norm denotes the energy amplification of the system. 

The observability Gramian of the system $\mat{P}\in\mathbb{R}^{2n \times 2n}$ can be used to formulate the $\mathcal{H}_2$ norm and therefore the optimal inertia allocation problem as \cite[Lemma 1]{poolla2017optimal}:
\begin{subequations}\label{eq:h2_norm_allocation}
\begin{align}
    \underset{\mathbf{P}, m_i}{\text{min}} \quad & \|\mathcal{G}\|_2^2=\mathrm{Trace} \, (\mathbf{B}^\top \mathbf{P} \, \mathbf{B}) \label{eq:objective_h2} \\
    \text{s.t.} \quad & \forall i \in \mathcal{N}, \nonumber \\
    &  \sum_{i\in\mathcal{N}}m_i \leq m_{\mathrm {budget}},\label{eq:totalinertiabudget_h2norm} \\
    & \underline{m}_{i} \leq m_{i} \leq \overline{m}_{i}, \label{eq:inertia_constraint_h2norm} \\
    & \mathbf{P} \mathbf{A} + \mathbf{A}^\top \mathbf{P} + \mathbf{C}^\top \mathbf{C} = 0, \label{eq:constraint3_h2} \\
    & \mathbf{P} \boldsymbol{z}_0 = \mathds{O}_{2n}\label{eq:constraint4_h2},
\end{align}
\end{subequations}
where the matrices $\mathbf{A}$ and $\mathbf{C}$ are defined in \eqref{eq:state} and \eqref{eq:h2_performance_output}, while the matrix $\mathbf{B}$ is modified from \eqref{eq:state} to consider impulse inputs at all the buses. The observability Gramian $\mathbf{P}$ is defined by the Lyapunov equation \eqref{eq:constraint3_h2} and subject to the constraint \eqref{eq:constraint4_h2} where $\mathds{O}_{2n}$ denotes the 2n dimensional vector of zeros.

\section{Numerical Results}\label{sec:res}

\subsection*{System Setup}
\begin{figure}[b!]
    \centering
     \vspace{-0.4cm}
     \includegraphics[width=1\linewidth]{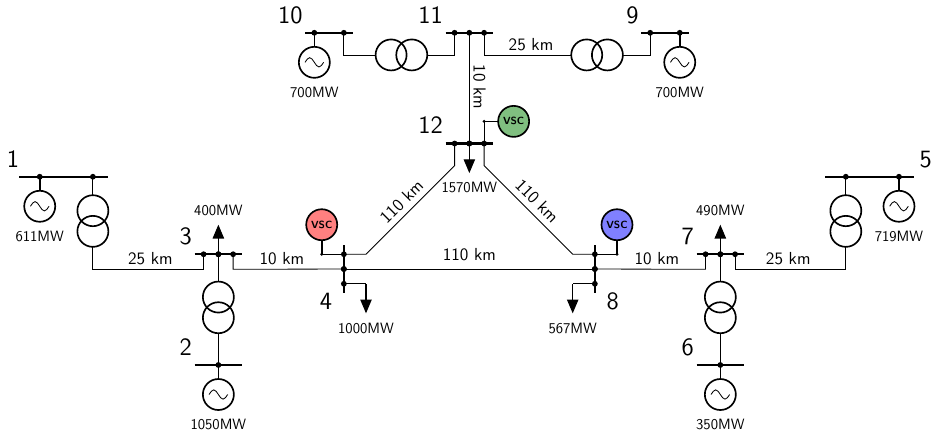}
    \caption{Topology of the 3-area test case with transformer impedance $z_{\mathrm{t}}=j \, 0.0167$ \rm{p.u.} and line impedance $z_{\mathrm{line}}=(0.0001 + j \, 0.001)$ \rm{p.u./km} (for $S_{\mathrm{b}}=100$ MW, $U_{\mathrm{b}}^{\mathrm{low}}=20$ kV for the three areas and $U_{\mathrm{b}}^{\mathrm{high}}=230$ kV for the transmission system connecting them).}
    \label{fig:3_area}
\end{figure}

In this section, the proposed dynamic optimization framework from \ref{subsec:dyn_opt} is tested on a 3-area system that is generally used for optimal inertia and damping allocation studies \cite{poolla2017optimal}, \cite{Ademola2018}. This system, shown in Fig.~\ref{fig:3_area}, is a modification of the well-known 2-area system from \cite{kundur1994power}, with an additional third area. We assume that \acp{VSC} are connected to buses $k \in \{4, 8, 12 \}$. Thus, the inertia and damping at these buses can be tuned in the proposed optimization framework while the parameters at the other \ac{SG} buses remain fixed. 

The frequency nadir is limited to a range between $0.98\,\rm{p.u.}$ and $1.02\,\rm{p.u.}$, according to \cite{ENTSOE2021Frequency}. Thus, we set $\dot{\theta}_{\rm{max/min}}=\pm0.02\, {\rm{p.u.}}$ Furthermore, the \ac{RoCoF} limits are set as $\dot{\omega}_{\rm{max/min}}=\pm0.02\, {\rm{p.u.}/s}$ in accordance with \cite{rocof2018}. The inertia and damping parameters for the non-tunable \ac{SG} buses are fixed as given in \cite{poolla2017optimal}, while for the tunable \ac{VSC} buses, they are lower bounded by the initial values, as provided in Table~\ref{tab:results}, and upper bounded by $\upperlim{m}_{j}=\upperlim{d}_{j}=120\,\rm{p.u.}$,\,$\forall{j}\in\mathcal{N}_{\rm{c}}$. The total allocated budget for inertia and damping is set as $m_{\rm{budget}}=d_{\rm{budget}}=250\,\rm{p.u.}$ Furthermore, we set $C_{\xi}$ at a very high value to ensure feasibility and prioritize energy dissipation and control effort equally by maintaining a 1:1 ratio between $C_{\kappa}$ and $C_{\vect{x}}$. Finally, we set $c_{d_j}=c_{m_j}=1,\,\forall j \in\mathcal{N}_{\rm c}$ in order to weigh the buses equally and balance inertia and damping. 

\subsection{Performance Analysis}

\begin{figure*}[h!]
    \centering
    \begin{tikzpicture}[scale=0.825]
    
    \draw[thick] (2.5,-1.7) rectangle (18,-2.3);
    \newcommand{\smallrectwidth}{1}

    \filldraw[fill=red, draw=black] (3, -1.875) rectangle (3+\smallrectwidth, -2.125); 
    \node at (5.55,-1.975) {Initial allocation};

    \filldraw[fill=blue, draw=black] (8, -1.875) rectangle (8+\smallrectwidth, -2.125); 
    \node at (10.8,-2.0) {$\mathcal{H}_2$ norm allocation};

    \filldraw[fill=darkgreen, draw=black] (13, -1.875) rectangle (13+\smallrectwidth, -2.125); 
    \node at (15.8,-2.025) {Dynamic allocation};

    \pgfplotsset{
        grid style={dashed,gray},
    }

    \begin{groupplot}[
        group style={
            group size=3 by 1,
            horizontal sep=1.75cm, 
        },
        ybar,
        grid=major,
        width=0.4\textwidth,
        height=5cm,
        clip=true, 
        tick align=inside 
    ]

    \nextgroupplot[
        ylabel={RoCoF (p.u./s)},
        xlabel={Dist. magnitude (p.u.)},
        symbolic x coords={Bus 5, Bus 8 (-1), Bus 8 (-3)},
        xtick=data,
        xticklabel style={align=center},
        xticklabels={Bus 5\\$-1$~p.u., Bus 8\\$-1$~p.u., Bus 8\\$-3$~p.u.},
        enlarge x limits=0.20,
        ymin=0.0, ymax=0.04,
        scaled y ticks = false,
        y tick label style={/pgf/number format/fixed,
            /pgf/number format/fixed zerofill,/pgf/number format/precision=2}
    ]
    
    \addplot[ybar,black,fill=red] coordinates {(Bus 5, 0.0105) (Bus 8 (-1), 0.0117) (Bus 8 (-3), 0.035)};
    \addplot[ybar,black,fill=blue] coordinates {(Bus 5, 0.0096) (Bus 8 (-1), 0.0027) (Bus 8 (-3), 0.0066)};
    \addplot[ybar,black,fill=darkgreen] coordinates {(Bus 5, 0.0094) (Bus 8 (-1), 0.0096) (Bus 8 (-3), 0.0079)};
    \addplot[black,line legend,sharp plot, nodes near coords={},shorten >=-8mm,shorten <=-8mm, update limits=false, line width= 1.5pt]coordinates {(Bus 5,0.02) (Bus 8 (-3),0.02)}node[midway,above left, xshift=-1.42em]{RoCoF limit};

    \nextgroupplot[
        ylabel={Frequency nadir (p.u.)},
        xlabel={Dist. magnitude (p.u.)},
        symbolic x coords={Bus 5, Bus 8 (-1), Bus 8 (-3)},
        xtick=data,
        xticklabel style={align=center},
        xticklabels={Bus 5\\$-1$~p.u., Bus 8\\$-1$~p.u., Bus 8\\$-3$~p.u.},
        enlarge x limits=0.20,
        ymin=0, ymax=0.12,
        ytick = {0, 0.02, 0.04, 0.06, 0.08, 0.10, 0.12},
        y tick label style={/pgf/number format/fixed,
            /pgf/number format/fixed zerofill,/pgf/number format/precision=2}
    ]
    
    \addplot[black,fill=red] coordinates {(Bus 5, 0.0356) (Bus 8 (-1), 0.0355) (Bus 8 (-3), 0.1066)};
    \addplot[black,fill=blue] coordinates {(Bus 5, 0.0085) (Bus 8 (-1), 0.0083) (Bus 8 (-3), 0.0121)};
    \addplot[black,fill=darkgreen] coordinates {(Bus 5, 0.0086) (Bus 8 (-1), 0.0083) (Bus 8 (-3), 0.0122)};
    \addplot[black,line legend,sharp plot, nodes near coords={},shorten >=-8mm,shorten <=-8mm, update limits=false, line width= 1.5pt]coordinates {(Bus 5,0.02) (Bus 8 (-3),0.02)}node[midway,above left, xshift=-1.5em]{nadir limit};

    \nextgroupplot[
        ylabel={Control effort (p.u.)},
        symbolic x coords={Bus 5, Bus 8 (-1), Bus 8 (-3)},
        xtick=data,
        xticklabel style={align=center},
        xticklabels={Bus 5\\$-1$~p.u., Bus 8\\$-1$~p.u., Bus 8\\$-3$~p.u.},
        enlarge x limits=0.20,
        ymin=0, ymax=600,
        ytick={0, 100, 200, 300, 400, 500, 600},
        xlabel={Dist. magnitude (p.u.)},
    ]
    \addplot[black,fill=red] coordinates {(Bus 5, 54.9196) (Bus 8 (-1), 54.9196) (Bus 8 (-3), 54.9196)};
    \addplot[black,fill=blue] coordinates {(Bus 5, 167.0860) (Bus 8 (-1), 273.3077) (Bus 8 (-3), 386.2067)};
    \addplot[black,fill=darkgreen] coordinates {(Bus 5, 147.6127) (Bus 8 (-1), 147.6173) (Bus 8 (-3), 358.6411)};
    \addplot[black,line legend,sharp plot, nodes near coords={},shorten >=-8mm,shorten <=-8mm, update limits=false, line width= 1.5pt]coordinates {(Bus 5,500) (Bus 8 (-3),500)}node[midway,above, xshift=0]{$m_{\text{budget}}+d_{\text{budget}}$};
    
    \end{groupplot}
    \end{tikzpicture}
    \vspace{-0.6cm}
    \caption{Grouped bar plots showing (left) RoCoF and (middle) frequency nadir for the disturbed bus, and (right) control effort ($\sum_{i \in \{4,8,12\}}(m_i+d_i)$) for the initial, $\mathcal{H}_2$ norm and dynamic optimization allocation for different disturbance locations and magnitudes.}
    \label{fig:performance_metrics}
     \vspace{-0.5cm}
\end{figure*}
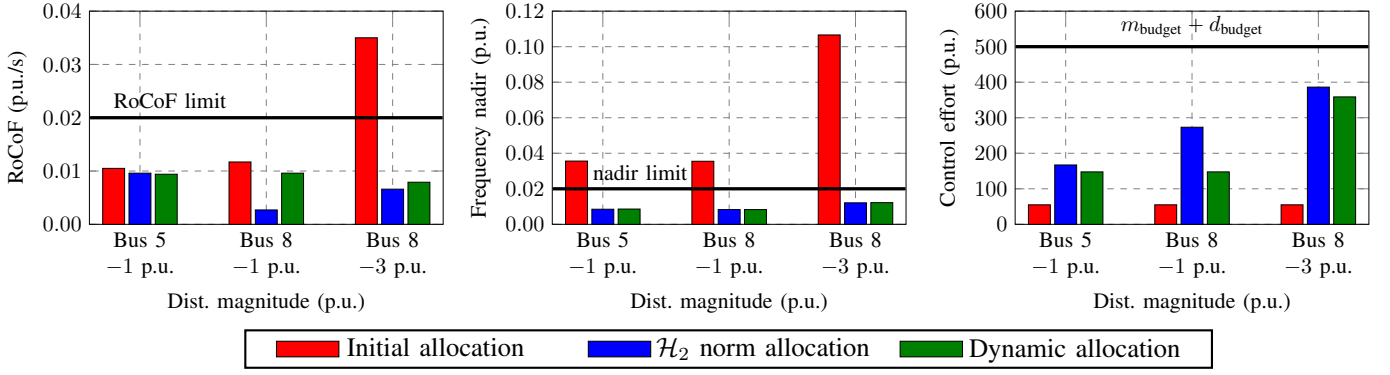

First, we compare the proposed dynamic optimization approach with the $\mathcal{H}_{2}$ norm-based optimal allocation from \cite{poolla2017optimal} summarized in Section~\ref{sec:H2norm_allocation}. Specifically, we consider a disturbance of $-1\,\rm{p.u.}$ at a \ac{SG} Bus~5 and disturbances of $-1\,\rm{p.u.}$ and $-3\,\rm{p.u.}$ at a \ac{VSC} Bus~8. We take into account the localized nature of the disturbance for both optimization approaches. Moreover, we optimize only the inertia at the \ac{VSC} buses for both approaches, since the $\mathcal{H}_{2}$ norm-based approach from \cite{poolla2017optimal} does not optimize the damping at the \ac{VSC} buses. Thus, to ensure a fair comparison between both approaches, we set the damping values in the system to the optimal values obtained from the dynamic optimization approach. We compare the frequency nadir and the \ac{RoCoF} for the disturbed bus, and the control effort, which is defined as the sum of inertia and damping at the tunable \ac{VSC} buses. The \ac{RoCoF} is measured using a moving average filter with a time window of 200 milliseconds as transients that vanish within this time window are irrelevant for \ac{RoCoF} relays~\cite{beyond_low_inertia}. On a personal computer equipped with 8 GB RAM and a 4-core i5-10300H processor, the dynamic optimization is completed in 3 seconds, while the $\mathcal{H}_2$ norm-based method completes in 1 second.

\begin{table}[b]
    \centering
    \vspace{-0.55cm}
    \caption{Initial and optimized values of inertia and damping for each of the tunable buses for the different analyzed scenarios.}
    \renewcommand\footnoterule{\vspace*{-5pt}}
    \begin{center}
    \scalebox{0.72}{%
        \begin{tabular}{c|c|c|c|c|c|c|c}
    \toprule
    Disturbance & Allocation & $m_4$ & $m_8$ & $m_{12}$ & $d_4$ & $d_8$ & $d_{12}$ \\
    \midrule
    / & Initial & 7.2193 & 12.7324 & 19.9899 & 2.7072 & 4.7746 & 7.4962 \\
    \midrule
    $-1$~p.u.& Dynamic Opt. & 7.2193 & 12.7324 & 19.9899 & 2.7072 & 97.4677 & 7.4962 \\
    at Bus 5& $\mathcal{H}_2$ norm & 7.2752 & 32.1295 & 20.0102 & 2.7072 & 97.4677 & 7.4962 \\
    \midrule
    $-1$~p.u.& Dynamic Opt. & 7.2193 & 12.7324 & 19.9899 & 2.7072 & 97.4721 & 7.4962 \\
    at Bus 8 & $\mathcal{H}_2$ norm & 25.6270 & 119.9807 & 20.0243 & 2.7072 & 97.4721 & 7.4962 \\
    \midrule
    $-3$~p.u.& Dynamic Opt. & 7.2193 & 92.6378 & 19.9899 & 111.2980 & 120.0000 & 7.4962 \\
    at Bus 8 & $\mathcal{H}_2$ norm & 7.4149 & 119.9897 & 20.0080 & 111.2980 & 120.0000 & 7.4962 \\
    \bottomrule
    \end{tabular}}
    \end{center}
    \label{tab:results}
\end{table}

The results of the optimal allocation for the methods are summarized in Table~\ref{tab:results}. The performance metrics for comparison are presented in Fig.~\ref{fig:performance_metrics}. From these results, we draw the following observations:

For the disturbance of $-1$~p.u. at the non-tunable \ac{SG} Bus~5, the proposed dynamic optimization only changes the damping at the tunable \ac{VSC} buses and does not change the inertia at these buses. In contrast, the $\mathcal{H}_2$ norm-based optimal allocation modifies the value of inertia at all the tunable buses, particularly at Bus~8 as it is the closest bus to the disturbed Bus~5. Thus, the proposed approach uses less control effort while also ensuring that the frequency nadir and \ac{RoCoF} are within their prescribed limits. Moreover, the \ac{RoCoF} is lower for the proposed allocation compared to the $\mathcal{H}_{2}$ norm-based allocation. 

For the disturbance of $-1$~p.u. at the tunable \ac{VSC} Bus~8, the dynamic optimization ensures that the frequency nadir and \ac{RoCoF} are within the prescribed limits by only increasing the damping at Bus~8 and not changing the inertia values. On the other hand, the $\mathcal{H}_2$ norm-based allocation increases the inertia at all the tunable \ac{VSC} buses, particularly at Bus~8, leading to a lower \ac{RoCoF} for a considerably higher control effort. The dynamic optimization formulation can directly constrain the state trajectories through constraints \eqref{eq:freq_constraint} and \eqref{eq:rocof_constraint}, keeping metrics within set limits with minimal control effort, unlike the $\mathcal{H}_2$ norm optimal allocation.

For the disturbance of $-3$~p.u. at the tunable \ac{VSC} Bus~8, the dynamic optimization results in a large control effort to ensure that the metrics are within the specified limits. The $\mathcal{H}_2$ norm-based allocation results in an even larger control effort for a marginally improved \ac{RoCoF}. For dynamic optimization, only the inertia at the disturbed bus is increased, and it correlates with the disturbance magnitude, i.e., higher inertia values for higher disturbance magnitudes. Consequently, lower values of \ac{RoCoF} can be attributed to inertia allocation in the system rather than an increase in overall system inertia. Contrary to this, the $\mathcal{H}_2$ norm algorithm allocates the inertia based on the network coherency, i.e., initial inertia and damping allocation and the Laplacian matrix.

With these observations, we note that the additional features in the proposed dynamic optimization, particularly the explicit constraints on the state trajectories, the weighted terms for control effort and energy dissipation, and the sensitivity to disturbance type, location, and magnitude, lead to improved results compared to the $\mathcal{H}_2$ norm-based allocation. 

\subsection{Sensitivity Analysis} \label{subsec:impact_dist}
Since dynamic optimization can allocate inertia and damping for specific locations and magnitudes of disturbances, we analyze how the optimal allocation changes with respect to the different disturbances. Regarding the disturbance location, we focus solely on the tunable \ac{VSC} buses, and we vary the magnitudes of the disturbance from $-3$~p.u. to 3 p.u. with a 0.25 p.u. resolution. 
The results obtained from the optimization algorithm for each magnitude and location of the disturbance are presented in Fig.~\ref{fig:disturbance_location_magnitude}, from which the following conclusions can be drawn:

\begin{figure}[b!]
    \centering
    \vspace{-0.4cm}
    \begin{tikzpicture}[scale=0.8]
        \begin{groupplot}[
            group style={
                group name=my plots,
                group size=2 by 3,
                xlabels at=edge bottom,
                ylabels at=edge left,
                horizontal sep=1.8cm,
                vertical sep=1.25cm 
            },
            width=0.28*\textwidth,
            height=4.95cm,
            grid=major,
            legend style={at={(1.25,-3.1)}, anchor=north, legend columns=3, /tikz/every even column/.append style={column sep=1cm}}, 
            legend image post style={xscale=1.5}, 
        ]
        ]
        \nextgroupplot[
            xlabel={Dist. magnitude (p.u.)},
            ylabel={$m_{i}^{opt}$ (p.u.)},
            grid=both,
            xmin=-3, xmax=3,
            ymax=130, ymin=0,
            ytick={0,30,60,90,120},
            xtick={-3,-2,-1,0,1,2,3},
        ]
        \addplot[mark=*, color=red] table [x index=0, y index=1, col sep=space] {data/disturbance_sensitivity/optimization_results.txt};
        \addlegendentry{$i=4$}
        \addplot[mark=*, color=blue] table [x index=0, y index=7, col sep=space] {data/disturbance_sensitivity/optimization_results.txt};
        \addlegendentry{$i=8$}
        \addplot[mark=*, color=darkgreen] table [x index=0, y index=13, col sep=space] {data/disturbance_sensitivity/optimization_results.txt};
        \addlegendentry{$i=12$}

        \nextgroupplot[
            xlabel={Dist. magnitude (p.u.)},
            ylabel={$d_{i}^{\text{ opt}}$ (p.u.)},
            grid=both,
            xmin=-3, xmax=3,
            ymax=130, ymin=0,
            ytick={0,30,60,90,120},
            xtick={-3,-2,-1,0,1,2,3},
        ]
        \addplot[mark=*, color=red] table [x index=0, y index=4, col sep=space] {data/disturbance_sensitivity/optimization_results.txt};
        \addlegendentry{d3\_4}
        \addplot[mark=*, color=blue] table [x index=0, y index=10, col sep=space] {data/disturbance_sensitivity/optimization_results.txt};
        \addlegendentry{d3\_8}
        \addplot[mark=*, color=darkgreen] table [x index=0, y index=16, col sep=space] {data/disturbance_sensitivity/optimization_results.txt};
        \addlegendentry{d3\_12}
        \legend{}

        \nextgroupplot[
            xlabel={Dist. magnitude (p.u.)},
            ylabel={$m_{i}^{\text{opt}}$ (p.u.)},
            grid=both,
            xmin=-3, xmax=3,
            ymax=130, ymin=0,
            ytick={0,30,60,90,120},
            xtick={-3,-2,-1,0,1,2,3},
        ]
        \addplot[mark=*, color=red] table [x index=0, y index=2, col sep=space] {data/disturbance_sensitivity/optimization_results.txt};
        \addlegendentry{m6\_4}
        \addplot[mark=*, color=blue] table [x index=0, y index=8, col sep=space] {data/disturbance_sensitivity/optimization_results.txt};
        \addlegendentry{m6\_8}
        \addplot[mark=*, color=darkgreen] table [x index=0, y index=14, col sep=space] {data/disturbance_sensitivity/optimization_results.txt};
        \addlegendentry{m6\_12}
        \legend{}

        \nextgroupplot[
            xlabel={Dist. magnitude (p.u.)},
            ylabel={$d_{i}^{\text{ opt}}$ (p.u.)},
            grid=both,
            xmin=-3, xmax=3,
            ymax=130, ymin=0,
            ytick={0,30,60,90,120},
            xtick={-3,-2,-1,0,1,2,3},
        ]
        \addplot[mark=*, color=red] table [x index=0, y index=5, col sep=space] {data/disturbance_sensitivity/optimization_results.txt};
        \addlegendentry{d6\_4}
        \addplot[mark=*, color=blue] table [x index=0, y index=11, col sep=space] {data/disturbance_sensitivity/optimization_results.txt};
        \addlegendentry{d6\_8}
        \addplot[mark=*, color=darkgreen] table [x index=0, y index=17, col sep=space] {data/disturbance_sensitivity/optimization_results.txt};
        \addlegendentry{d6\_12}
        \legend{}

        \nextgroupplot[
            xlabel={Dist. magnitude (p.u.)},
            ylabel={$m_{i}^{\text{opt}}$ (p.u.)},
            grid=both,
            xmin=-3, xmax=3,
            ymax=130, ymin=0,
            ytick={0,30,60,90,120},
            xtick={-3,-2,-1,0,1,2,3},
        ]
        \addplot[mark=*, color=red] table [x index=0, y index=3, col sep=space] {data/disturbance_sensitivity/optimization_results.txt};
        \addlegendentry{m9\_4}
        \addplot[mark=*, color=blue] table [x index=0, y index=9, col sep=space] {data/disturbance_sensitivity/optimization_results.txt};
        \addlegendentry{m9\_8}
        \addplot[mark=*, color=darkgreen] table [x index=0, y index=15, col sep=space] {data/disturbance_sensitivity/optimization_results.txt};
        \addlegendentry{m9\_12}
        \legend{}

        \nextgroupplot[
            xlabel={Dist. magnitude (p.u.)},
            ylabel={$d_{i}^{\text{ opt}}$ (p.u.)},
            grid=both,
            xmin=-3, xmax=3,
            ymax=130, ymin=0,
            ytick={0,30,60,90,120},
            xtick={-3,-2,-1,0,1,2,3},
        ]
        \addplot[mark=*, color=red] table [x index=0, y index=6, col sep=space] {data/disturbance_sensitivity/optimization_results.txt};
        \addlegendentry{d9\_4}
        \addplot[mark=*, color=blue] table [x index=0, y index=12, col sep=space] {data/disturbance_sensitivity/optimization_results.txt};
        \addlegendentry{d9\_8}
        \addplot[mark=*, color=darkgreen] table [x index=0, y index=18, col sep=space] {data/disturbance_sensitivity/optimization_results.txt};
        \addlegendentry{d9\_12}
        \legend{}
        \end{groupplot}
    \end{tikzpicture}
     \vspace{-0.2cm}
    \caption{Optimal values of inertia $m_{i}$ and damping $d_{i}$ at each tunable bus $i\in\mathcal{N}_{\rm c}$ for disturbance at Bus~4 (top), Bus~8 (middle), and Bus~12 (bottom).}
    \label{fig:disturbance_location_magnitude}
\end{figure}
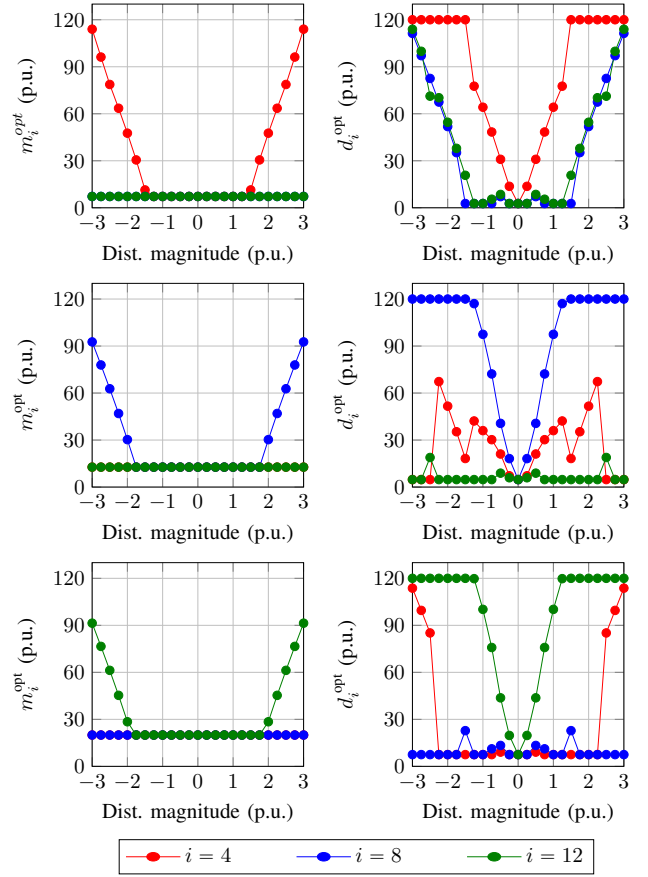

\begin{enumerate}
    \item[1)] Inertia allocation is affected solely by local disturbances. Specifically, the inertia budget for a particular bus is used to address disturbances at that particular bus. This dependence of inertia allocation on disturbance location is also observed in \cite{poolla2017optimal}.
    \item[2)] On the other hand, the damping allocation at a certain bus is determined by disturbances at that bus and disturbances at neighboring buses. For larger disturbance magnitudes, damping at the non-disturbed buses is used to provide sufficient overall damping for the system.
    \item[3)] For low but increasing disturbance magnitudes, the proposed dynamic optimization algorithm increases the damping at the tunable buses and, as the magnitude of the disturbance increases further, the inertia is additionally increased at the tunable buses.
\end{enumerate}

\begin{figure}[t!]
    \centering
    \begin{tikzpicture}[scale=0.80]
        \begin{axis}[
            xlabel={Dist. magnitude $j$ (p.u.)},
            ylabel={Probability $\lambda_{i,j}$},
            width={0.6\textwidth},
            height=4.7cm,
            grid=major,
            xmin=-3, xmax=3,
            ymin=-0.002, ymax=0.045,
            legend style={
                at={(0.5,1.0)},
                anchor=north
            },
            scaled y ticks=false,
            yticklabel style={
                /pgf/number format/.cd,
                fixed,
                precision=3
            }
        ]
        \addplot[color=red, mark=*, smooth, tension=0.5] 
        table [x index=0, y index=1, col sep=space] {data/disturbance_sensitivity/disturbances_and_probabilities_3.txt};
        \addlegendentry{$i=4$}

        \addplot[color=blue, mark=*, smooth, tension=0.5] 
        table [x index=0, y index=1, col sep=space] {data/disturbance_sensitivity/disturbances_and_probabilities_6.txt};
        \addlegendentry{$i=8$}

        \addplot[color=darkgreen, mark=*, smooth, tension=0.5] 
        table [x index=0, y index=1, col sep=space] {data/disturbance_sensitivity/disturbances_and_probabilities_9.txt};
        \addlegendentry{$i=12$}

        \end{axis}
    \end{tikzpicture}
    \vspace{-0.7cm}
    \caption{Disturbance probabilities for the tunable VSC buses $i$.}
    \label{fig:probabilities}
    \vspace{-0.40cm}
\end{figure}
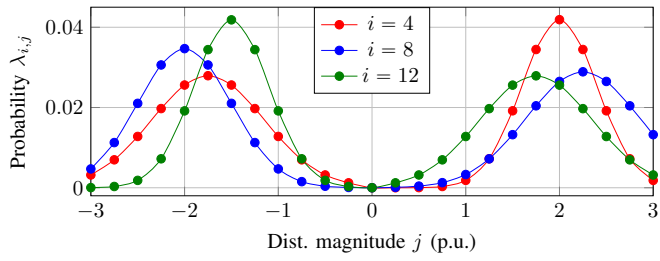

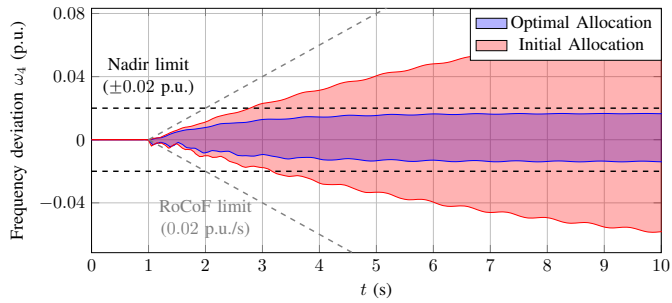
\begin{figure}[h!]
    \centering
    \vspace{-0.0cm}
    \resizebox{0.49\textwidth}{!}{\begin{tikzpicture}[scale=1.0]
        \begin{axis}[
            width=0.7\textwidth,
            height=0.35\textwidth,
            grid=both,
            xlabel={$t$ (s)},
            ylabel={Frequency deviation $\omega_4$ (p.u.)},
            xmin=0, xmax=10,
            ytick={-0.12, -0.08, -0.04, 0, 0.04, 0.08, 0.12},
            scaled y ticks = false,
            yticklabel style={/pgf/number format/.cd,fixed,precision=3},
            legend style={at={(0.8325, 1.0)}, anchor=north},
        ]
        
        \addplot[name path=opt_small, mark=none, color=blue, forget plot] 
            table [x index=0, y index=1, col sep=space, skip first n=4] 
            {data/disturbance_sensitivity/time_domain_results/bus3_min_nadir_optimized.txt};
            
        \addplot[name path=opt_large, mark=none, color=blue, forget plot] 
            table [x index=0, y index=1, col sep=space, skip first n=4] 
            {data/disturbance_sensitivity/time_domain_results/bus3_max_nadir_optimized.txt};
            
        \addplot[name path=init_small, mark=none, color=red, forget plot] 
            table [x index=0, y index=1, col sep=space, skip first n=4] 
            {data/disturbance_sensitivity/time_domain_results/bus3_min_nadir_initial.txt};
            
        \addplot[name path=init_large, mark=none, color=red, forget plot] 
            table [x index=0, y index=1, col sep=space, skip first n=4] 
            {data/disturbance_sensitivity/time_domain_results/bus3_max_nadir_initial.txt};
            
        \addplot[fill=blue, fill opacity=0.3] fill between[of=opt_small and opt_large];
        \addlegendentry{Optimal Allocation}
        
        \addplot[fill=red, fill opacity=0.3] fill between[of=init_small and init_large];
        \addlegendentry{Initial Allocation}

        \draw[thick, dashed, black] (axis cs:0, 0.02) -- (axis cs:30, 0.02);
        \draw[thick, dashed, black] (axis cs:0, -0.02) -- (axis cs:30, -0.02);

        \draw[thick, dashed, gray!100] 
            (axis cs:1, 0) -- (axis cs:40, {0 + 0.02 * (40 - 1)});
        \draw[thick, dashed, gray!100] 
            (axis cs:1, 0) -- (axis cs:40, {0 - 0.02 * (40 - 1)});

       \node[black, align=center] at (axis cs:1, 0.04) {Nadir limit \\ ($\pm 0.02$ p.u.)};
       \node[gray!100, align=center] at (axis cs:2, -0.05) {RoCoF limit \\ ($0.02$ p.u./s)};
       
    \end{axis}
    \end{tikzpicture}}
    \vspace{-0.70cm}
    \caption{Comparison of initial model and optimized model frequency trajectories at Bus~4 for various disturbance locations and magnitudes obtained from Monte Carlo simulations.}
    \label{fig:probabilities_results}
     \vspace{-0.0cm}
\end{figure}

Past disturbance data collected by system operators could be used to obtain the probabilities of disturbances occurring in the power system. As an example, we assume that the disturbances considered occur with probability $\lambda_{i,j}$ as shown in Fig.~\ref{fig:probabilities}, where $i\in\{4,8,12\}$ and $j\in[-3,3]$. We can therefore determine the optimal allocation of inertia and damping for the system based on these probabilities using the following weighted average approach
\begin{equation}
        x_k^{\text{opt}} = \frac{\sum_{i\,\in\{4,8,12\}} \sum_{j=-3}^{3} \lambda_{i,j} \, x_{k,i,j}^{\text{opt}}}{\sum_{i\,\in\{4,8,12\}} \sum_{j=-3}^{3} \lambda_{i,j}},
    \label{eq:prob}
\end{equation}
where $x\in\{m,d\}$. For evaluation purposes, we perform 100 Monte Carlo simulations of the system using the value obtained from \eqref{eq:prob} and with the disturbance drawn from the probability distributions given in Fig.~\ref{fig:probabilities}. The result is illustrated in Fig.~\ref{fig:probabilities_results}, where we compare the envelope of the obtained frequency trajectories at Bus~4. We observe that the optimal allocation ensures that the frequency nadir and \ac{RoCoF} constraints are respected in all cases, which is not the case for the initial allocation.

\section{Conclusion} \label{sec:concl}
This paper introduces a novel formulation based on dynamic optimization to allocate inertia and damping at \ac{VSC} buses to maintain frequency stability. This method minimizes post-fault energy loss and control effort in the form of bus inertia and damping. Moreover, it can directly constrain the post-fault state trajectories and consider various types of disturbances for the optimal allocation. In the presented case studies, we first show that the proposed method achieves allocations that utilize the control effort more efficiently while maintaining frequency stability compared with an $\mathcal{H}_2$ norm-based allocation for different disturbances. We also demonstrate that inertia allocation at a bus is impacted by local disturbances, whereas damping allocation can be affected by either local disturbances or disturbances in nearby buses. Finally, we validate a weighted average approach to allocate inertia and damping considering various disturbances and their probabilities. 

\bibliographystyle{IEEEtran}
\bibliography{bibliography.bib}

\end{document}